# Small Signal Stability Analysis of Kurdistan Regional Power System

**Ibrahim Ismael Hamarash, PhD**
Department of Electrical Engineering
Salahaddin University-Erbil
Kurdistan, Iraq
<ibrahim.hamarash@su.edu.iq>

### Abstract

This paper presents for the first time a mathematical model for evaluating the Planned Kurdistan Regional Power System (KRPS) for its ability to maintain stability under small disturbances and fluctuations during normal operating conditions. To achieve this objective, practical field data, manufacture's datasheets, related IEEE task force reports have been used to build a complete mathematical model in MATLAB®/SIMULINK®/SimPowerSystem® environment. New modules have been established and added to the platform wherever it does not support special type of elements. The model represents accurately all the power system components involved in physical phenomena of system dynamic oscillations. The model consists of 53 transmission lines, 35 nodes and 6 generating stations. The system is simulated under different configurations and settings; the dynamic behaviors associated with each configuration are recorded and analyzed accordingly.

## INTRODUCTION

It is well known that electrical power systems consist of thousands of individual components that interact together, giving rise to various electromechanical modes of oscillations even during normal operating conditions. If these oscillations continue without sufficient damping, they may lead to system instability. Practical definitions of power system security generally require sufficient damping of oscillatory modes. The main way in which oscillations can arise is spontaneous oscillations due to gradual change in system operating conditions [1].

The seriousness of power system oscillations lie on running of the system close to operating condition limitations. In recent years, power transactions are increasing in volume and variety in restructured power systems because of the large amounts of money to be made in exploiting geographic differences in power prices and costs. Restructured power systems are expected to be operated at a greater variety of operating points and closer to their operating constraints. The onset of low frequency inter-area oscillation is one operational constraint which already limits bulk power transactions under some conditions [2].

Spontaneous oscillations arise when the mode damping becomes negative by a gradual change in system operating conditions. The oscillations grow and may reach a steady state in which the oscillations persist at constant amplitude, or continue and lead to uncontrolled amplitude and serious system malfunctions.

Electromechanical oscillations are of the types; inter plant, local, inter area, control and tensional modes [2]. The frequency of oscillations and the phenomena behind each oscillation type is given in table 1.

**Table 1 Low Frequency Oscillation Categories and Their Frequencies**

| No | Oscillation Type | Frequency of Oscillation | Oscillation Phenomena |
|----|------------------|--------------------------|------------------------|
| 1 | Inter-Plant | 2.0-3.0 Hz | Machines of the same power station site oscillate against each other. The rest of the system is unaffected. |
| 2 | Local-Plant | 1.0-2.0 Hz | One generator swings against the rest of the system. The impact of the oscillation is localized to the generator and the line connecting it to the grid . The rest of the system is normally modeled as a constant voltage source whose frequency is assumed to remain constant. |
| 3 | Inter-Area Oscillation | 0.3-1.0 Hz | Oscillations of real power flow between regions of a power system or groups of generators. This phenomenon is observed over a large part of the network. The variation in tie-line power can be large. The damping characteristic of the inter area mode is dictated by the tie-line strength, the nature of the loads and the power flow through the interconnection and the interaction of loads with the dynamics of generators and their associated controls. |
| 4 | Control-Mode Oscillation | 4-10 Hz | These Oscillations are associated with generators and poorly tuned exciters, governors, HVDC converters and SVC controls. Loads and excitation systems can interact through control modes.Transformer tap-changing controls can also interact in a complex manner with non-linear loads giving rise to voltage oscillations |
| 5 | Torsional Oscillation | 10-46 Hz | Torsional oscillations involving interaction of generator shaft modes with the power system. These modes are associated with a turbine generator shaft system in the frequency range of 10-46 Hz. Usually these modes are excited when a multi-stage turbine generator is connected to the grid system through a series compensated line. A mechanical torsional mode of the shaft system interacts with the series capacitor at the natural frequency of the electrical network. The shaft resonance appears when network natural frequency equals synchronous frequency minus torsional frequency. |

During the history of power industry, inadequate damping of the above mentioned oscillations have led to many system separations and blackout, for example: Detroit Edison (DE-Ontario Hydro (OH)-Hydro Quebec (HQ) (1960s, 1985), Finland-Sweden-Norway-Denmark (1960s),Saskatchewan-Manitoba Hydro Western Ontario (1966),Italy-Yugoslavia-Austria (1971-1974), Western Electric Coordinating Council (WECC) (1964,1996), Mid-continent area power pool (MAPP) (197 1,1972), South East Australia (1975), Scotland-

England (1978), Western Australia (1982,1983), Taiwan(1985), Ghana-Ivory Coast (1985), Southern Brazil (1975-1980,1984) [3,4,5].

In this study, for the first time, a model has been build for Kurdistan Regional Power System (KRPS)/Iraq for stability studies. The components; synchronous generators with their excitation and governing systems, transmission lines, and loads are modeled. Real data from the system are used when available, otherwise, manufacturer's datasheets and/or related IEEE task force reports and standards are used instead. The system is simulated under various operating conditions, and then the results have been analyzed from stability point of view.

## MODELING OF KURDISTAN REGIONAL POWER SYSTEM
### i. System Description

Kurdistan Regional Power System(KRPS) comprises two hydro power stations, Dokan (400 MVA) and Darbandikhan (249 MVA) in addition to two gas stations cited at Perdawood (500 MVA) and Chamchamal (750 MVA). Another two more gas stations are under construction in Duhok (200 MVA) and TaqTaq(200 MVA). The four gas stations are private sector investments. Currently, KRPS uses 35 load nodes at 132 kV substation ends totaling 2202 MW. The system serves five million population in an area of 80,000 square km[6,7].

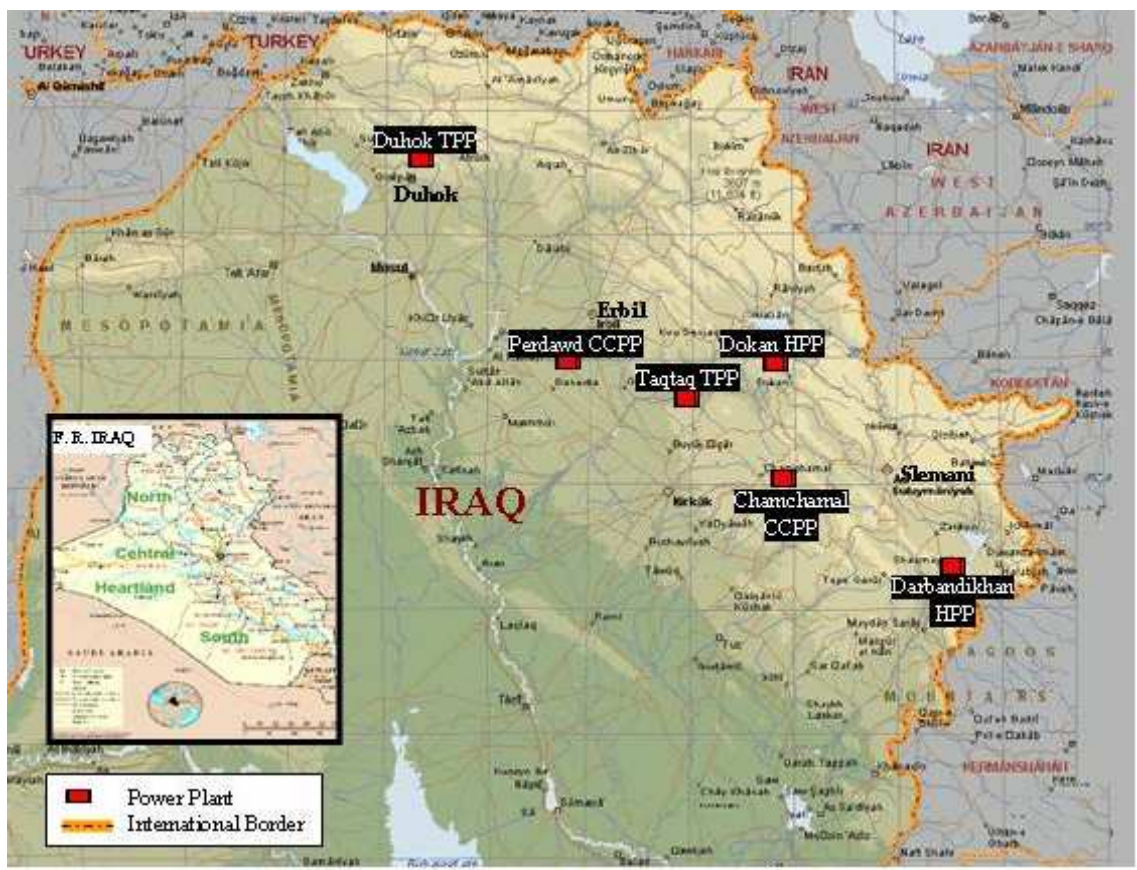

Fig.1 Map of the Investigated Power System of Kurdistan Region/ Iraq

### ii. KRPS Modeling

Power system models are often conveniently defined in terms of the major subsystems of equipment. Only these dynamics of the equipment which are active in determining the system performance need to be considered. The following components are important to the dynamic study: the synchronous machines, the excitation systems, the prime movers and governors, Transmission Lines, FACTS & SVC devices (if applicable) and power system loads.

The dynamic behavior of these devices are described through a set of differential equations but the power flow in the network is represented by a set of algebraic equations. This gives rise to a set of differential-algebraic equations (DAE) representing the power system mathematical mode [8].

**a. Synchronous Generators**

The dynamic behavior of the generators within a power system is of fundamental importance to the overall quality of the power supply. The synchronous generator converts mechanical power to electrical power at a specific voltage and frequency. The source of the mechanical power, the prime mover, may be a diesel engine, a steam turbine, a gas or a water turbine. Whatever the source, it must have the basic property that its speed is almost constant regardless of the power demand. The analysis of any power system to determine its stability involves the mechanical properties of the machines because, after any disturbance, they must adjust the angle of their rotors to meet the conditions of power transfer imposed. The electric dynamics have very short time constants compared to hydrodynamics and can be ignored [9]. The general approach to synchronous machine modeling is quite mature. Their Mathematical models vary from elementary classical models to more detailed ones. It depends on the nature of the study. For low frequency oscillation studies, the sub transient phenomena have to be captured. We have considered a d-q axis modeling of all synchronous generators in this study (Dukan, Derbandikhan, Perdawad and Chamchamal, TaqTaq and Duhok stations) using IEEE conventions.

Consider an interconnected power system with m-machines and n-buses. We consider four windings on the rotor (one field and one damper in d-axis and two dampers in q axis). For i=1 to m, the following equations represent machine dynamics [10].

$$\frac{d\delta_i}{\delta t} = \omega_i - \omega_s \tag{1}$$

$$\frac{d\omega_i}{dt} = \frac{\omega_s}{2H}[T_{mi} - D(\omega_i - \omega_s) - \frac{X''_{di} - X_{lsi}}{X'_{di} - X_{lsi}} E'_{qi} I_{qi} - \frac{X'_{di} - X''_{di}}{X'_{di} - X_{lsi}} \psi_{1di} I_{qi}$$
$$- \frac{X''_{qi} - X_{lsi}}{X'_{qi} - X_{lsi}} E'_{di} I_{di} + \frac{X'_{qi} - X''_{qi}}{X'_{qi} - X_{lsi}} \psi_{2qi} I_{di} + (X''_{qi} - X''_{di}) I_{qi} I_{di}] \tag{2}$$

$$\frac{dE'_{qi}}{dt} = \frac{1}{T'_{doi}}[-E'_{qi} - (X_{di} - X'_{di})\{-I_{di} - \frac{(X'_{di} - X''_{di})}{(X'_{di} - X'_{lsi})^2}$$
$$(\psi_{1di} - (X'_{di} - X_{lsi})I_{di} - E'_{qi})\} + E_{fdi}] \tag{3}$$

$$\frac{dE'_{di}}{dt} = \frac{1}{T'_{qoi}}[-E'_{di} - (X_{qi} - X'_{qi})\{-I_{qi} - \frac{(X'_{qi} - X''_{qi})}{(X'_{qi} - X'_{lsi})^2}$$
$$(\psi_{2qi} + (X'_{qi} - X_{lsi})I_{qi} - E'_{di})\}] \tag{4}$$

$$\frac{d\psi_{1di}}{dt} = \frac{1}{T''_{doi}}[-\psi_{1di} + E'_{qi} + (X'_{di} - X_{lsi})I_{di}] \tag{5}$$

$$\frac{d\psi_{2qi}}{dt} = \frac{1}{T''_{qo}}[-\psi_{2qi} + E'_{di} + (X'_{qi} - X_{lsi})I_{qi}] \tag{6}$$

Where for *ith machine*

| | |
|---|---|
| m : | total number of generators, |
| $\delta$ : | generator rotor angle(rad), |
| $\omega$ : | rotor angular speed(rad per second), |
| $E_{qi}'$ : | transient emf due to field flux-linkage(p.u), |
| $E_{di}'$ : | transient emf due to flux-linkage in q-axis damper coil(p.u), |
| $\psi_{1di}$ : | sub-transient emf due to flux-linkage in d-axis damper(p.u), |
| $\psi_{2qi}$ : | sub-transient emf due to flux-linkage in q-axis damper(p.u), |
| $I_{di}$ : | d-axis component of stator current(p.u), |
| $I_{qi}$ : | q-axis component of stator current(p.u), |
| $X_{di}, X_{di}', X_{di}''$ : | synchronous, transient and sub-transient reactances(p.u), respectively along d-axis, |
| $X_{qi}, X_{qi}', X_{qi}''$ : | synchronous, transient and sub-transient reactances(p.u), respectively along q-axis, |
| $T_{do}', T_{do}''$ : | d-axis open-circuit transient and sub-transient time constants(second), respectively |
| $T_{qo}', T_{qo}''$ : | q-axis open-circuit transient and sub-transient time constants(second), respectively |

The subscripts d and q are for d and q axis quantity, R and s for Rotor and stator quantity, l and m for Leakage and magnetizing inductance, and f and k are for Field and damper winding quantity respectively.

**b. Excitation Systems**

All KRPS generating stations are equipped with fast IEEE ST1A excitation system. This type of excitation system is often modeled as a single time-constant block. The error signal is used as input and $E_{fd}$ as output. figure 2 shows a block diagram of IEEE ST1A[11] as proposed by IEEE task force for excitation system modeling. Because of the SimPowerSystems toolkit does not support this type of excitation system, a new model has been built using SIMULINK. The complete model is shown in figure 3.

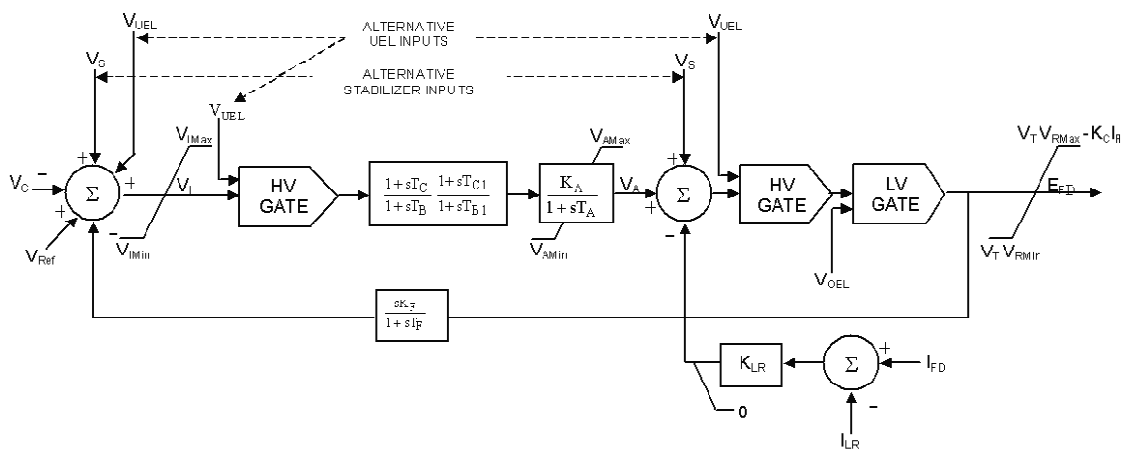

**Fig.2 Type IEEE ST1A Excitation System Block Diagram**

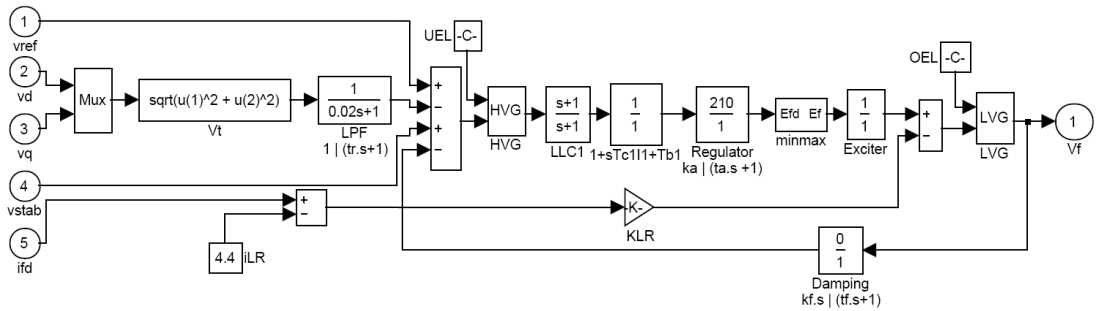

**Fig.3 SIMULINK® version of IEEEST1A Excitation System**

### c. Turbine System

KRPS comprises two types of turbine systems: Hydraulic for Dukan and Derbandikhan hydro stations, Gas Turbine for Perdawd , Chamchamal, Duhok and Taqtaq gas stations. The Hydraulic Turbine and Governor uses a nonlinear hydraulic turbine model, a PID governor system, and a servomotor [12,13], the block set of the hydraulic turbine with governing system is shown in figure 4.

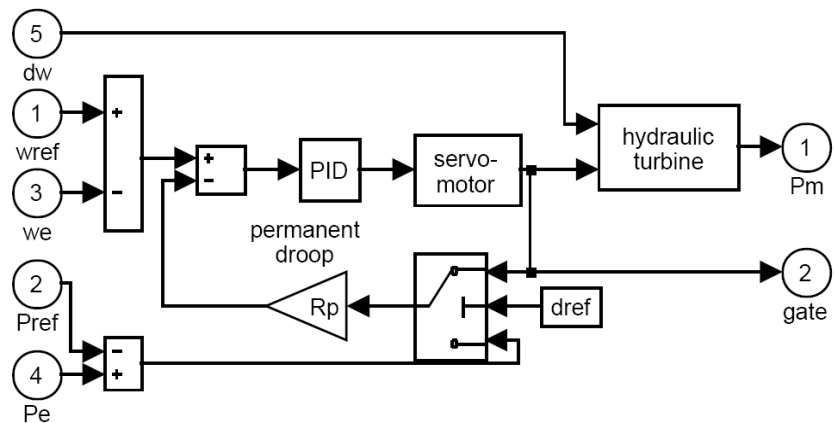

**Fig.4 The Hydraulic Turbine and Governing System**

Several research groups proposed several dynamic models of turbine-governors with varying degrees of complexity to represent different makes and models of gas turbine units[14,15,16]. A typical combined cycle plant configuration is proposed by IEEE working group on prime over and energy supply models for system dynamics in power system studies [17]. the configuration of this proposed system is shown in figure 5. This arrangement is made up of an air compressor, combustor, gas turbine   heat recovery boiler, and a steam turbine.

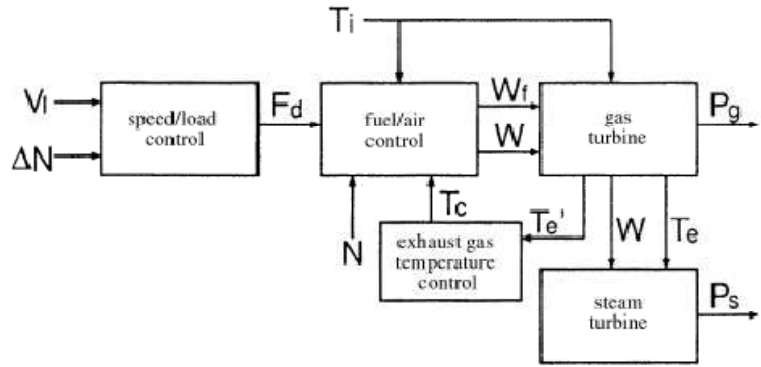

**Fig.5 Functional Block Diagram of CCGS**

A new model for the Gas governing system is developed in this study and realized in SIMULINK. The block set of the developed model is shown in figure 6. This model has been used in modeling all gas stations in the system.

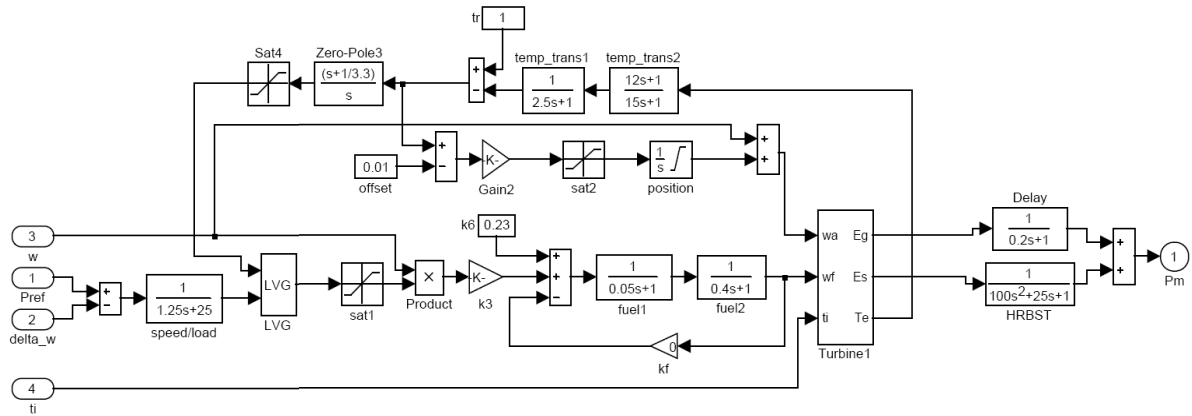

**Fig.6 The Developed Model for Gas Stations**

The turbine block in figure 6 is based on the well known heat cycle of the gas turbine which is shown in figure 7 [18].

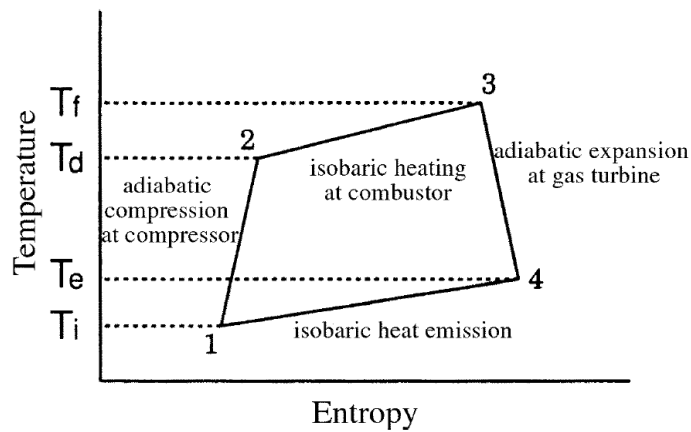

**Fig.7 Gas Turbine Heat Cycle**

**d. Load Models**

The modeling of loads in stability studies is a complex problem due to the unclear nature of aggregated loads (e.g. a mix of fluorescent, compact fluorescent, incandescent lamps, refrigerators, heater, motor, etc.). Static load model is a model that expresses the active and reactive powers at any instant of time as functions of the bus voltage magnitude and frequency at the same instant. Static load models are used both for essentially static load components, e.g., resistive and lighting load, and as an approximation for dynamic load components, e.g., motor-driven loads.

A static load model that represents the power relationship to voltage as an exponential equation, usually in the following form[19]:

$$P = P_o (\frac{v}{v_o})^a \tag{7}$$

$$Q = Q_o (\frac{v}{v_o})^b \tag{8}$$

where $P_o$ and $Q_o$ are the real and reactive powers consumed at a reference voltage $V_o$ respectively. While load modeling is done by specifying proportions of constant power, constant current, and constant impedance, the exponents a and b depend on the type of load that is being represented, e.g. for constant power load models a = b = 0, for constant current load models a = b = 1 and for constant impedance load models a = b = 2.

**e. Network Models**

Power in KRPS is transmitted over long distances through 132 kV overhead lines. It is usual that overhead lines are classified according to their lengths. In this study, lines shorter than 25 km are represented as equivalent three phase PI sections. The line parameters R, L, and C are specified as positive- and zero sequence parameters that take into account the inductive and capacitive couplings between the three phase conductors as well as the ground parameters. This method of specifying line parameters assumes that the three phases are balanced. For lines longer than 25 km, the distributed effects of the parameters are considered.

**f. The Complete Block Diagram Model for KRPS**

All subsystem models derived in previous sections for the power system components are integrated to establish a complete block diagram model of Kurdistan Regional Power System. The complete MATLAB®/SIMULINK®/ SIMPOWERSYSTEMS® based model of the system is shown in Appendix I. This model is used to study the small signal stability of KRPS in the following sections.

**SIMULATION**

A series of tests has been performed for the system for small signal disturbances and normal load fluctuations. During the tests, Dokan generating station has been selected as a swing bus and other stations as PV buses. The results of the Load Flow has been updated automatically for some blocks and manually for others. The solver ***ode23tb(stiff/TR-BDF2*** with a maximum step size of 1/50 has been used for solving the system equations.

In order to study the behavior of the base system, a time domain simulation and eigenvalue computation were performed for a 5% change in mechanical power input to one of the largest generating stations named Perdawood close to the largest load centre at Erbil city , the capital. Thus, Figure 8 shows the eigenvalues map on the complex plane, whereas figure 9 shows the corresponding load angle for all generation stations. From these two figures, one can conclude that the system is oscillatory, critically stable, poorly damped and it is very convenient to develop spontaneous oscillations. After 60 seconds the system yet oscillates critically. Any small disturbance may lead the system to loose synchronism.

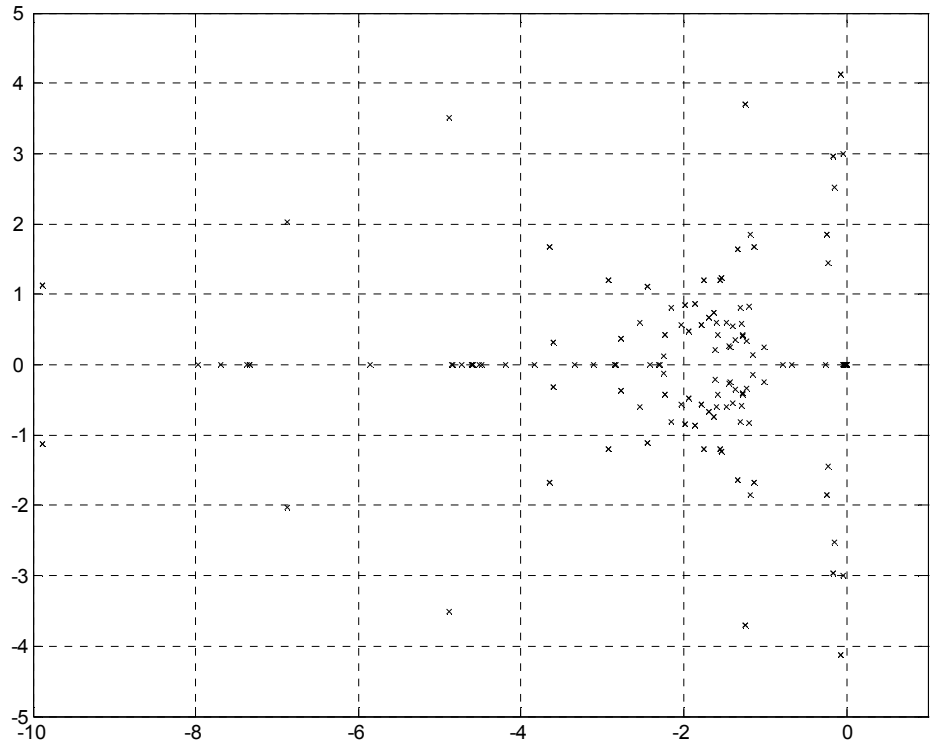

**Fig.8 Eignvalue map for the base system**

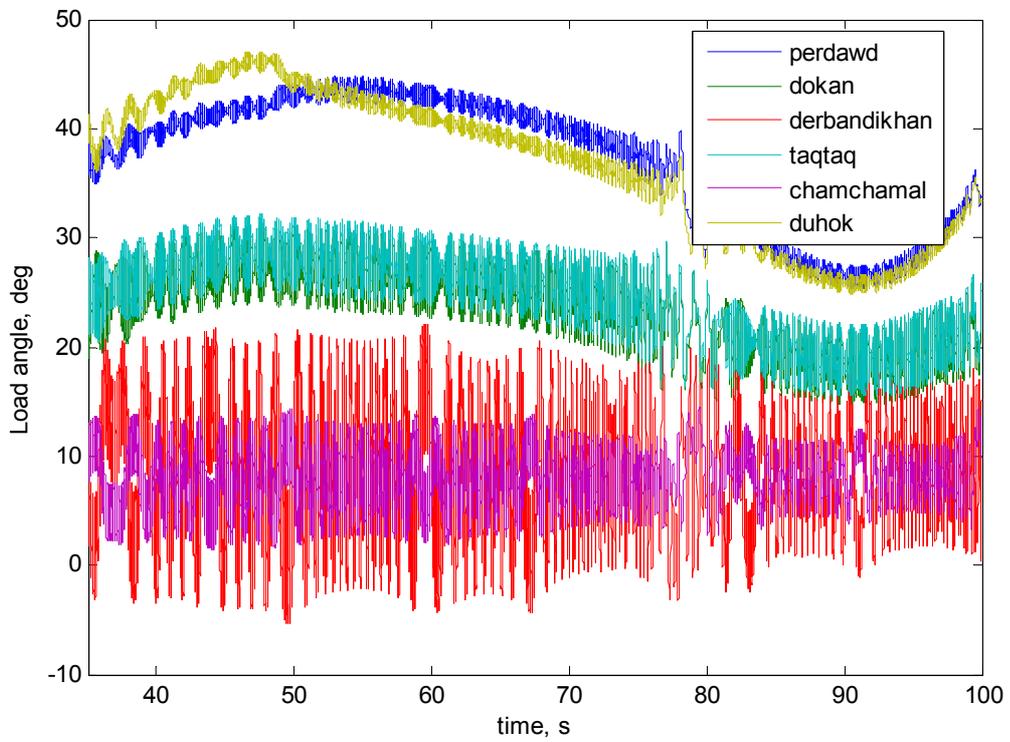

**Fig. 9 Load angle of all generation stations, without PSS**

To evaluate the dynamic performance of the system with PSS controller, a PSS has been added to the excitation sub-system. An IEEE type 4b, multi-band PSS is used in our study. This PSS is based on speed deviation as an input to produce output signal [20,21]. Simulation studies are carried out for the system subjected to small disturbance with and without system stabilizers.

In order to verify the effectiveness of PSS under small disturbance, the mechanical power input to Perdawood generation station is increased by 5% p.u at t = 40 sec [22]. The system response under this small disturbance for Perdawood station is shown in figures10-13, while the load angle and terminal voltage variations due to this disturbance for all generation stations are shown in figures 20 and 21 respectively. It is clear form the figures 14 and 15, that the PSS based controller has good damping characteristics to low frequency oscillations and quickly stabilizes the system under this small disturbance.

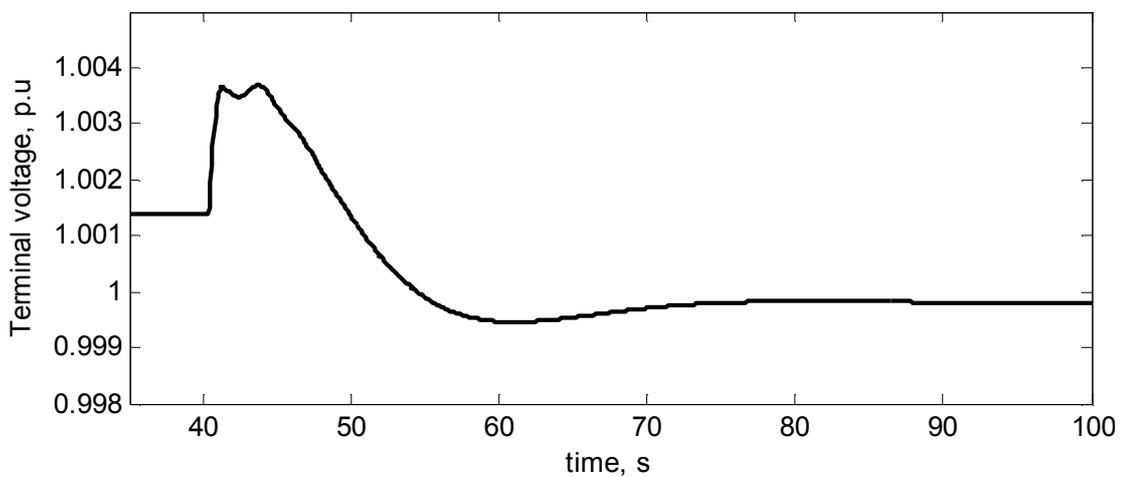

**Fig. 10 System Performance, Terminal Voltage of Perdawood, with PSS**

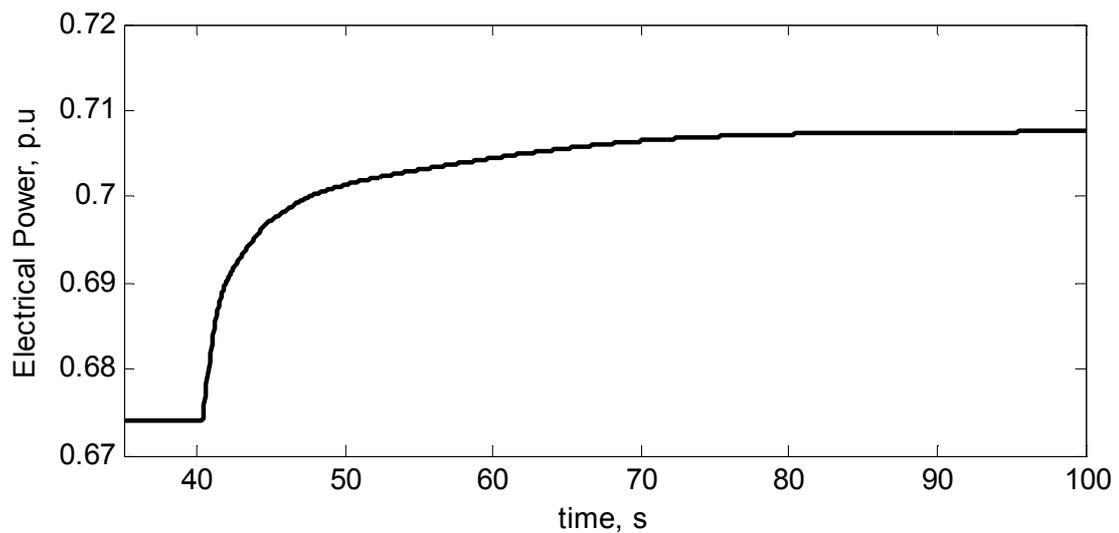

**Fig. 11 System Performance , Electrical power of Perdawppd, with PSS**

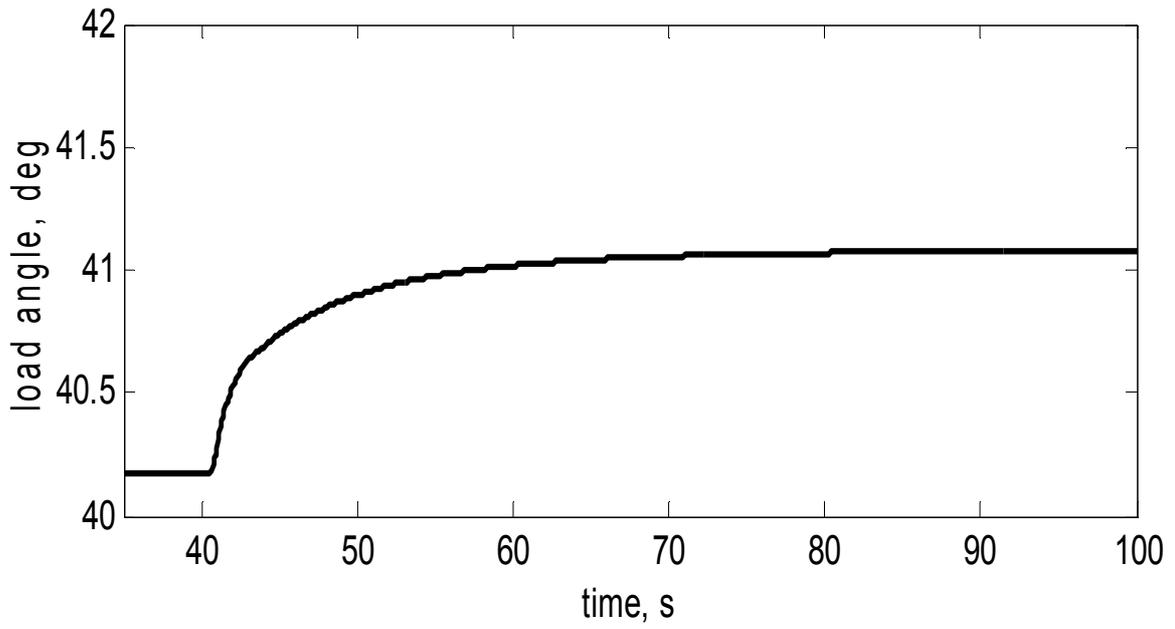

**Fig. 12 System Response (Load Angle) to Small Disturbance in Perdawood**

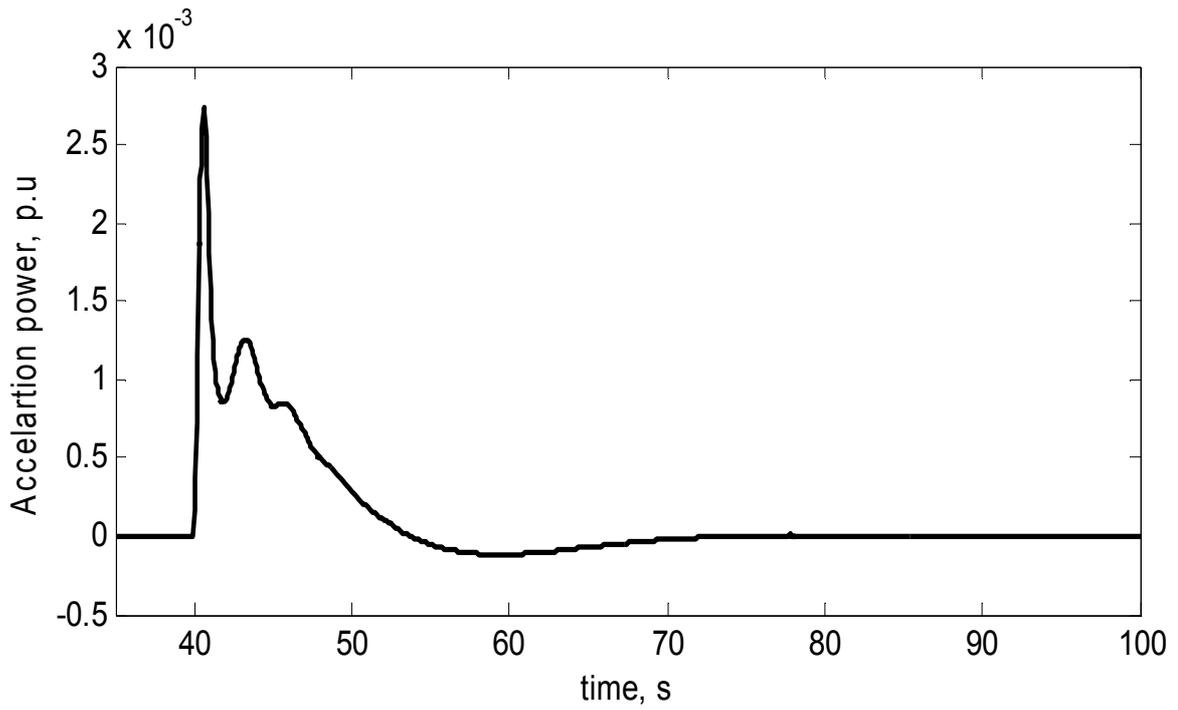

**Fig. 13 System Performance, Acceleration Power in Perdawood,**

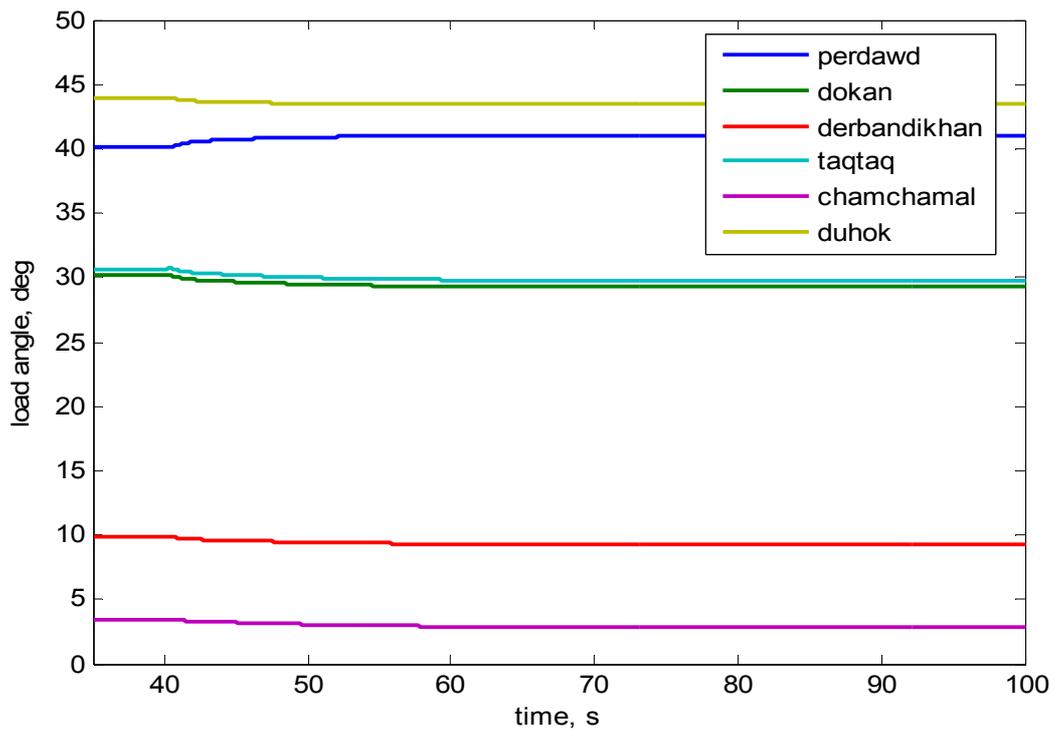

Fig. 14  System Response (Load Angle) to Small Disturbance in Perdawd

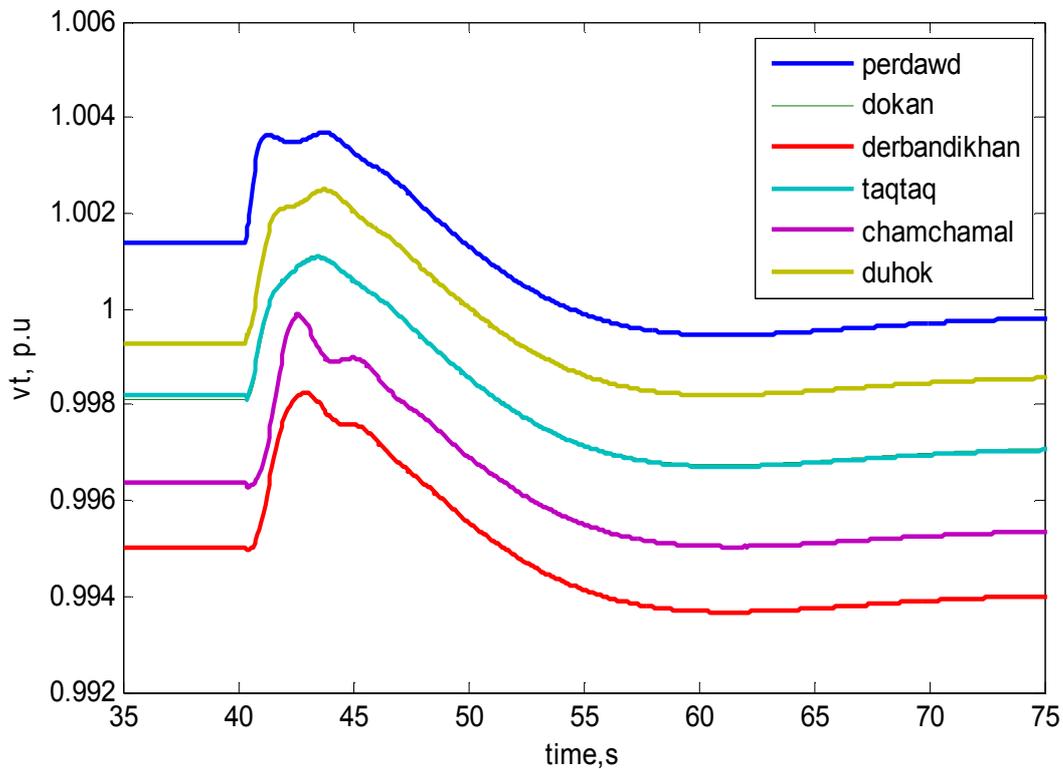

Fig.15  System Response (Terminal Voltage) to Small Disturbance in Perdawd

## CONCLUSIONS

This is the first stability analysis for Kurdistan Regional Power System. However, KRPS logically seems to be part of Iraqi national grid, but physically it is not, it operates as an isolated power system. In this situation, stability is a real concern and has to be studied from all possible points of view. The first step in this studying is the availability of reasonable mathematical model for the system. Hence, in this study, a complete mathematical model has been built. The components datasheets, daily operational reports, IEEE typical data and standards have been used in the modeling process to confirm its closeness to the real system. Each subsystem has been verified using IEEE benchmarks and standards, and then they have been used in the integrated system which has been used in the study of KRPS.

To test the small signal stability and the system's ability to restore its initial condition after occurrence of small disturbances, the eignvalue analysis method is used. The state matrix is isolated with its eignvalues spectrum shown in figure 8. The eignvalue map shows the oscillatory characteristic of the system, while most of the eignvalues are located close to the imaginary axis. The base system is critically stable, there are poor damped oscillations anywhere in the system, that could be overcome by implementing the suitable PSS controllers.

Our results show that the planned electricity system for Kurdistan Region is tightly meshed and electrically short, with the relative impedances between nodes quite small. Therefore, during a major contingency involving the loss of significant generation, the system will remain in synchronism, and the frequency deviation will be very similar at all points on the system.

Finally, this model is available free in charge at Salahaddin University-Erbil, it can be ordered from the researcher to be used as a bench mark, for more development, further study and future planing of the system.

**ACKNOWLEDGMENT**

This research is a part of an Endeavour Postdoctoral Research sponsored by Endeavour Research Fellowships Program of the Department of Education, Science and Training, Australia, under Endeavour ID: ERIRAQ_PDR_153_2007. The research is carried out in the Department of Electrical & Computer Systems Engineering, Faculty of Engineering, Monash University, Australia.

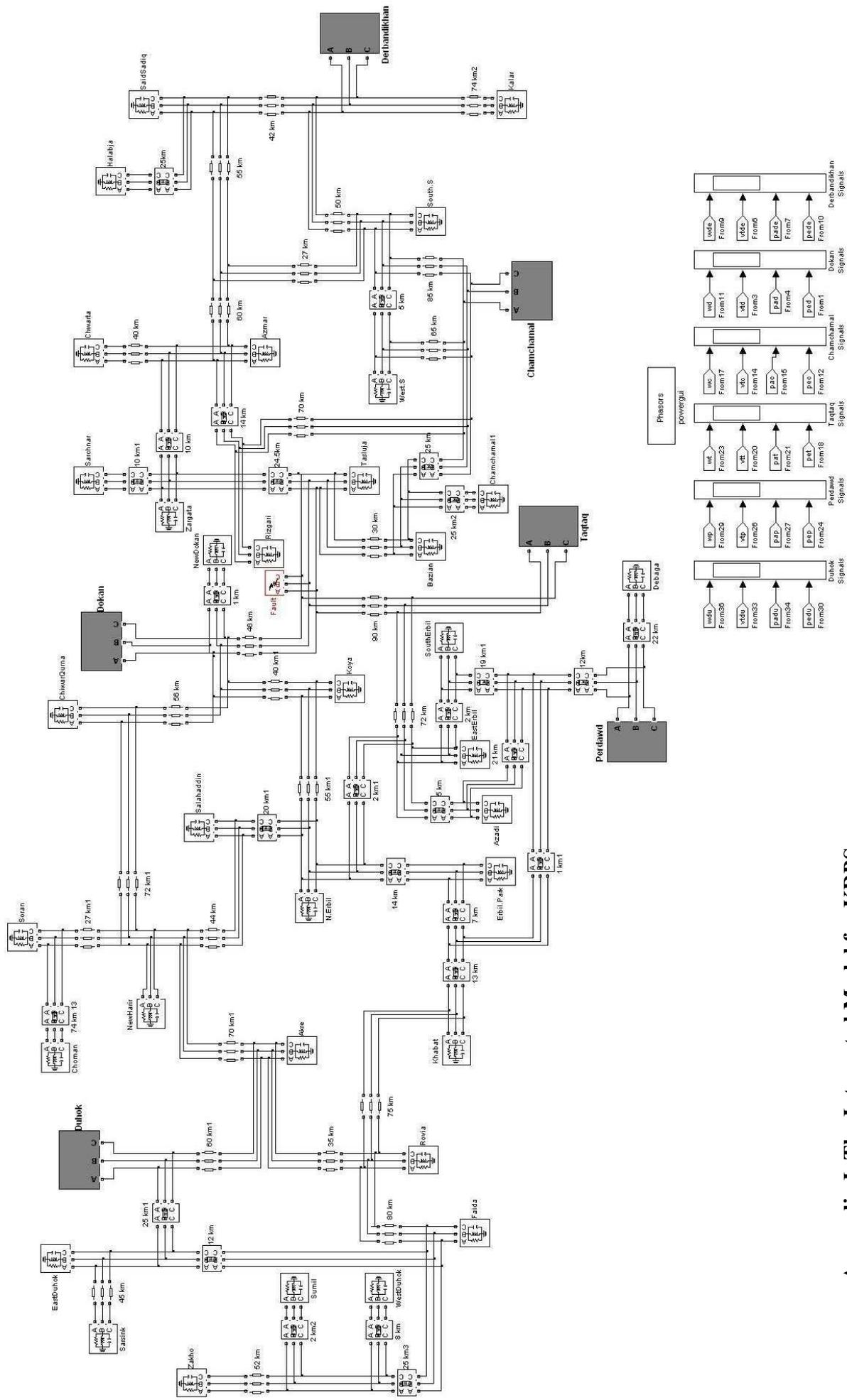

**Appendix I. The Integrated Model for KRPS**